\documentclass[twocolumn,floatfix,prl,showpacs,color,superscriptaddress,epsfig]{revtex4}
\usepackage{graphicx}
\usepackage{epsfig}
\usepackage{amsmath}
\usepackage{epstopdf}
\usepackage{bbold}
\begin{document}
\title{Metacinnabar ($\beta$-HgS): a strong 3D topological insulator with 
highly anisotropic surface states}
\author{Fran\c{c}ois Virot}
\affiliation{Institut de Mat\'eriaux, Micro\'electronique et Nanosciences de Provence, Facult\'e St. J\'er\^ome, Universit\'e Aix-Marseille, Case 142, F-13397 Marseille Cedex 20, France}
\author{Roland Hayn}
\affiliation{Institut de Mat\'eriaux, Micro\'electronique et Nanosciences de Provence, Facult\'e St. J\'er\^ome, Universit\'e Aix-Marseille, Case 142, F-13397 Marseille Cedex 20, France}
\author{Manuel Richter}
\affiliation{IFW Dresden, P.O. Box 270 116, D-01171 Dresden, Germany}
\author{Jeroen van den Brink}
\affiliation{IFW Dresden, P.O. Box 270 116, D-01171 Dresden, Germany}

\date{\today}

\begin{abstract}
We establish the presence of topologically protected edge states on the (001) surface of HgS in the zinc-blende structure using density-functional electronic structure calculations.  The Dirac point of the edge state cone is very close to the bulk valence band maximum. The Dirac cone is extremely anisotropic with a very large electron velocity along one diagonal of the surface elementary cell $x'$ and a nearly flat dispersion in the perpendicular direction $y'$. The strong anisotropy 
originates from a broken fourfold rotoinversion symmetry at the surface.
\end{abstract}

\pacs{}

\maketitle

Motivated by the ongoing search for spintronics materials, recent theoretical work on the effect of spin-orbit interactions on the band structure of solids predicted the existence of novel topological insulators (TI's) in two~\cite{Kane05a,Kane05b,Bernevig06} and three dimensions~\cite{Fu07,Fu07b,Moore07,Hasan10}. Topological insulators distinguish themselves from ordinary insulators due to the presence of a non-trivial topological invariant in the bulk band structure~\cite{Fu07,Fu07b,Moore07,Hasan10,Qi08}. Although there is an excitation gap in the bulk as in conventional band insulators, the topological bulk properties dictate that there be gapless modes at the interface between a TI and a topologically trivial state, for instance the vacuum. The surface states of a three-dimensional (3D) TI form massless Dirac cones, with a single Dirac cone being the simplest case. The potential applicability for spintronics arises because the surface states of such a 3D TI appear in time-reversed pairs in which the two electrons have both opposite spin and velocity. 

From the material's viewpoint, two-dimensional (2D) TI  behavior was experimentally observed in HgTe quantum wells~\cite{Konig07}, soon after the theoretical prediction~\cite{Bernevig06}. 3D TI's discovered so far are the binary compounds Bi$_2$Se$_3$, Sb$_2$Te$_3$, Sb$_2$Te$_3$~\cite{Zhang09,Xia09,Chen09} and the alloy Bi$_{1-x}$Sb$_x$~\cite{Fu07,Hsieh08} and very recently, 3D-TI behavior was proposed in certain ternary Heusler compounds~\cite{Chadov10,Lin10}. In the former the presence of Dirac cones has been established by photo-emission, but direct electronic transport evidence for  metallic surface states on top of an insulating bulk is lacking, mainly because of the small bulk gaps.
It is interesting to note that before the advent of TI's, in semi-metallic bismuth surface states were identified that are in hindsight very similar to the chiral surface states of TI's~\cite{Koroteev04}. 

Here we establish that metacinnabar, the binary compound $\beta$-HgS,  is a strong 3D topological insulator.  As opposed to other TI's discovered to date, the topologically protected Dirac cone on the (001) surface of HgS in the zinc-blende structure is extremely anisotropic.  
This originates from a broken fourfold rotoinversion symmetry at the surface. The presence of strongly anisotropic 
topologically protected modes implies highly one-directional spin and electron transport properties providing
a natural route to create potentially ideal quantum wires. 
As this electronic anisotropy is absent in the bulk because it is fourfold rotoinversion invariant, the presence of significant spin and charge transport anisotropy at the (001)  surface of HgS will be a smoking gun for its TI nature.

Metacinnabar, $\beta$-HgS, is a gray-black natural mineral, found in mercury deposits that are formed near-surface, under low-temperature conditions. The bulk electronic structure of HgS in the zinc-blende structure has been considered before, with contradicting conclusion on the existence of a band gap inversion~\cite{Delin02,Moon06,Cardona09}.  Delin~\cite{Delin02} and Cardona {\em et al}~\cite{Cardona09} find HgS to be insulating with an unconventional band order at the $\Gamma$ point: the conduction band minimum has $p$-character and the $s$-band, which usually builds the conduction band minimum, is about one eV below the Fermi level.  However, according to the electronic structure calculation of  Moon and Wei~\cite{Moon06}, such a band inversion due to spin-orbit interaction is absent. For lattice structures with inversion symmetry the parity criteria of Fu {\em et al}~\cite{Fu07b} guarantee that an inverted semiconductor is a topological insulator, but owing to the lack of inversion symmetry of the zinc-blende crystal structure of $\beta$-HgS these criteria do not apply. We will therefore establish the TI nature of $\beta$-HgS via the bulk-boundary correspondence and explicitly calculate its surface states.

To this end we performed high precision, all electron, full potential and fully relativistic electronic structure calculations for $\beta$-HgS using the FPLO  (full potential local orbital) code, version 9.01-35 ~\cite{koepernik,FPL}. We unambiguously establish the presence of an inverted band gap in the bulk material. Investigating in detail the (001) Hg-terminated surface in a slab geometry we demonstrate the presence of a highly anisotropic Dirac cone. This proves that  HgS is a topological insulator.  The anisotropy of the cone implies that the velocity and mobility of the charge carriers at the surface is very high in one direction and low in the direction perpendicular to it. Thus it will strongly affect surface spin and charge transport.

\begin{figure}
\begin{center}
\includegraphics[scale=0.55]{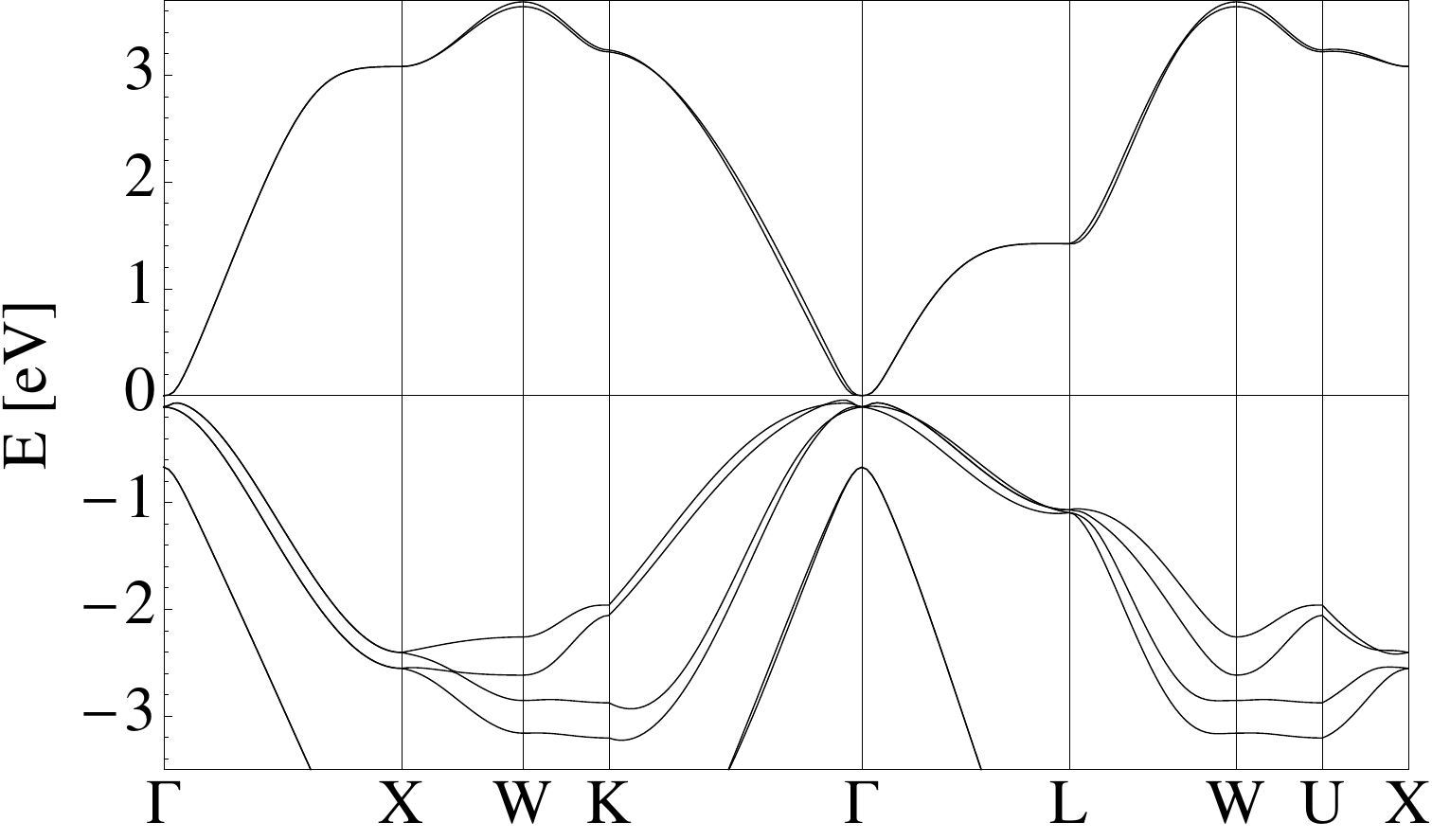}
\end{center}
\caption{Band structure of HgS in the bulk. It is an inverted semiconductor which is visible 
at the $\Gamma$-point: small gap and electron-like dispersion at the top of the valence band.} 
\label{f1}
\end{figure}

To calculate the bulk band structure of the zinc-blende crystal we used the experimental 
lattice parameter $a_0=5.85$ \AA,  incidentally equal to the theoretical value (Fig.~\ref{f1}). 
Without spin-orbit (SO) coupling, all three 
$3p$-orbitals of S would be degenerate leading to metallic behavior. The SO-coupling 
splits the $3p$-manifold into a lower quartet ($3p_{3/2}$) and an upper doublet 
($3p_{1/2}$) with a gap of 109 meV at the $\Gamma$-point. The indirect gap
between conduction and valence band is 42 meV. 
The band of $s$-character (here Hg $6s$)
which builds the conduction band minimum in ordinary semiconductors is found at 0.65 eV 
below the Fermi level. This level inversion, in connection with the electron like dispersion
of the valence band at the $\Gamma$-point implies that HgS is an inverted 
semiconductor, which is a prerequisite for it to be a topological insulator.

The insulating character of HgS is related to the fact that the sulfur quartet ($3p_{3/2}$) is 
below the doublet ($3p_{1/2}$) which contradicts the third Hund's rule for pure
 $p$-orbitals.
The related compounds HgSe and HgTe, in which the Hg $6s$ level is far below 
the Fermi level too, have the "correct order'' with the doublet below the quartet.  
Accordingly, there is no gap in those two compounds.
The reversed SO-order in HgS (quartet below doublet) comes about because of a small but significant contribution of Hg $5d$ orbitals whose SO-coupling dominates the one of S $3p$ and reverses the sign. So, contrary to naive expectations it is actually the {\it smallness} of the SO-coupling parameter of the sulfur $p$ orbitals (in contrast to the $p$ orbitals of Se or Te)~\cite{Delin02} which allows HgS to become a topological insulator.

\begin{figure}
\begin{center}
\includegraphics[scale=0.65]{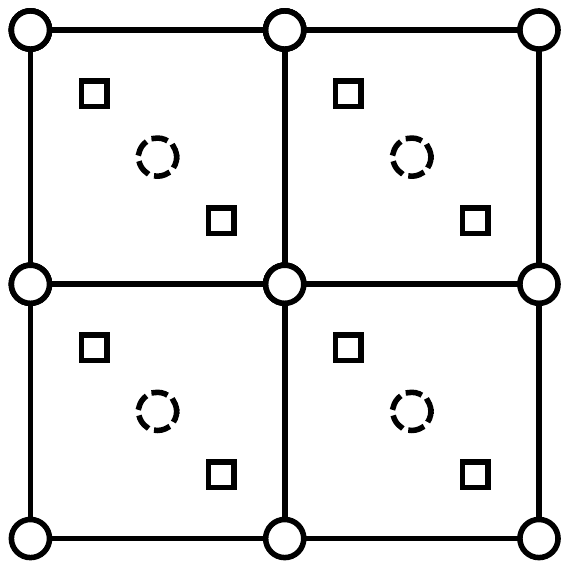}
\end{center}
\caption{Schematic representation of the (001) surface of HgS. Shown are the mercury top layer (full circles: Hg atoms, 
dashed circle: surplus Hg atom that has been removed) and the sub-surface sulfur layer (squares: S).  At the surface the rotoinversion symmetry of the bulk crystal structure is broken.} 
\label{f2}
\end{figure}

Since HgS has an inverted level order at the $\Gamma$-point, its topological invariant can be non-trivial and thus be different from the vacuum. This implies a change of topology at any surface and hence the presence of gapless surface states. An explicit calculation of
the topological invariant of HgS is technically 
more involved 
as the simple definition for lattices with inversion symmetry cannot be applied for zinc-blende
crystals which lack this symmetry~\cite{Xiao10}. Therefore, we studied in detail the (001) surface of $\beta$-HgS and
calculated its electronic structure explicitly, 
establishing the topological structure of the bulk via the bulk-boundary correspondence~\cite{Hasan10,Qi08}.
To this purpose we performed electronic structure calculations of HgS slabs, varying the thickness. To guarantee electro-neutrality of the surface each second Hg-atom has to be removed from a Hg-terminated surface (see Fig.~\ref{f2}). The surplus Hg-atoms in the top and bottom layers were removed in a checker-board pattern, so that the slab has tetragonal symmetry.  Such a surface reconstruction is known from similar surfaces like for instance the (100) surface of WO$_3$ \cite{Oliver96}.
We have also considered S-terminated surfaces but they are less relevant to the present discussion as we find such a termination to be 265 meV per surface elementary cell higher in energy and therefore substantially less stable.

The HgS slab that we consider has a 
tetragonal supercell (space group 111) consisting of (001) stacks of
conventional cubic elementary cells, spaced by an empty (vacuum) layer
of 23.4 \AA \ thickness. The experimental cubic lattice parameter, $a_0=5.85$ \AA,
was used as in the bulk calculations. Self-consistent calculations were performed in 
a scalar-relativistic mode, using the local density approximation in the
parameterization of Ref. \onlinecite{perdew92}.
The presented band structures were evaluated in the fully relativistic mode
(four-component Dirac-Kohn-Sham scheme) using a dense $k$-point mesh
of $25\times25\times1$ intervals in the full Brillouin zone and linear
tetrahedron integration with Bl\"ochl corrections in order to ensure the correct position of the
Fermi level. The valence basis set comprised sulfur (2s, 2p, 3s, 3p, 3d,
4s, 4p) and mercury (5s, 5p, 5d, 6s, 6p, 6d, 7s) states. 

\begin{figure}
\begin{center}
\includegraphics[scale=0.60]{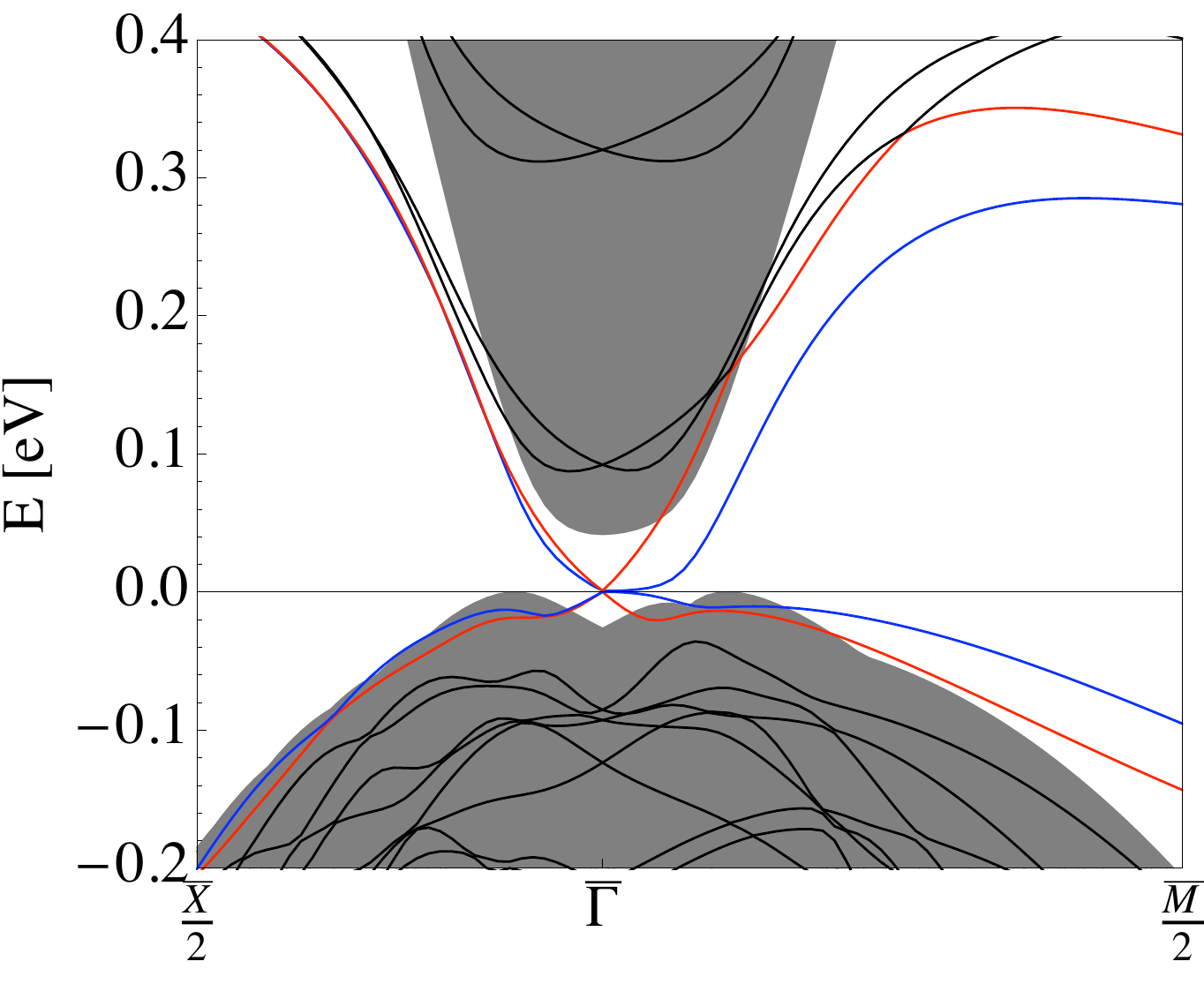}
\end{center}
\caption{(Color) Band structure of the 8-layer slab (lines) superimposed with the bulk band structure projected onto the surface (shaded area). The relevant surface bands are indicated in  red (top surface) and blue (bottom surface).}
\label{f3}
\end{figure}

\begin{figure}
\begin{center}
\includegraphics[scale=0.4]{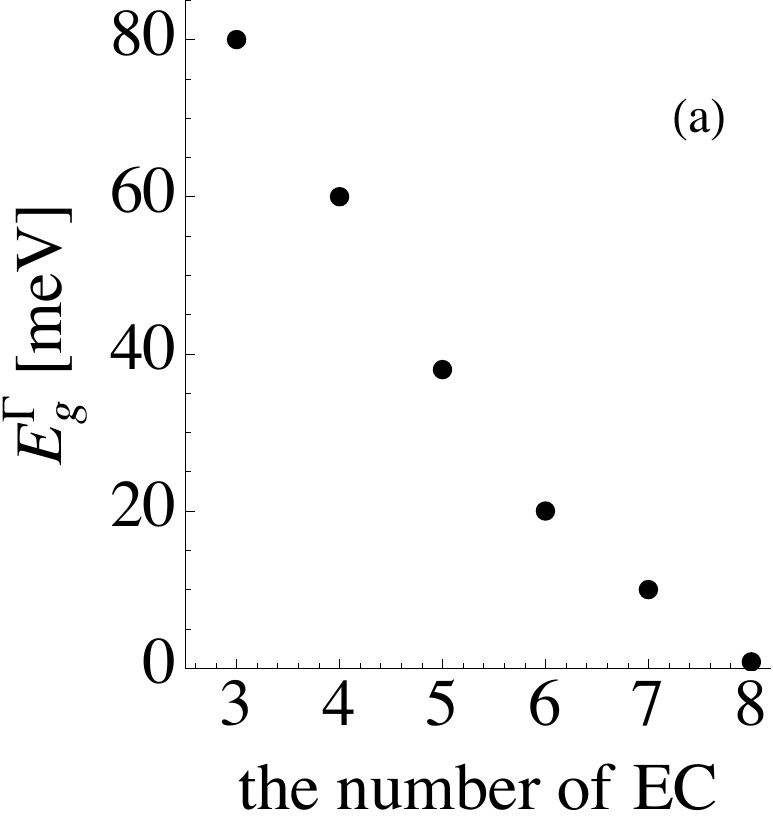}
\includegraphics[scale=0.4]{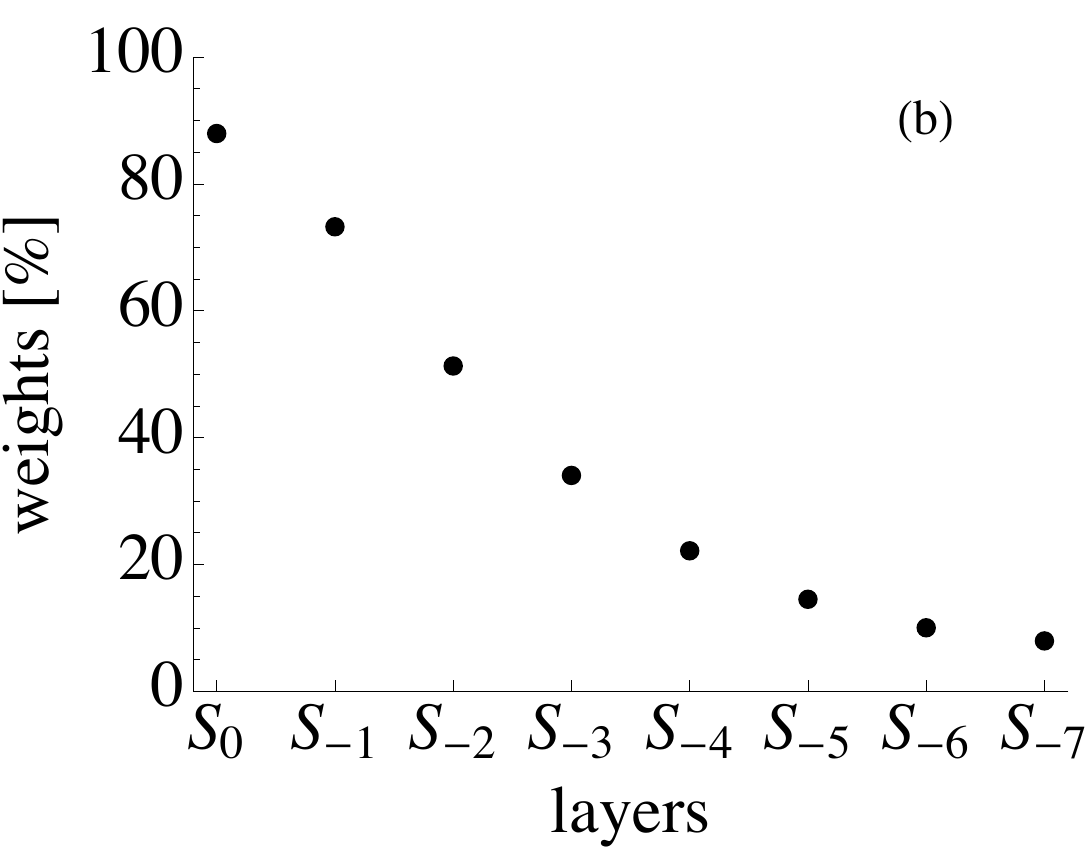}
\end{center}
\caption{(a) Width of the gap as function of the number of elementary cells (EC) and (b) sum of  3$p$ weights over the four surface states  at $\Gamma$, resolved for each sulfur layer in the slab of 8 EC.}
\label{f4}
\end{figure}

We systematically studied slabs containing $n$ elementary cells with $n$ ranging from 3 to 8. With increasing slab thickness the gap caused by the coupling of the surface states at opposite sides of the stack gradually closes (see Fig.~\ref{f4} a) and a band crossing appears which is most clearly seen for the 8-layer slab the band structure of which is shown in Fig.~\ref{f3}. In this figure the band structure of the slab calculation is superimposed with the bulk band structure projected onto the surface. There clearly are four bands crossing the Fermi level at the $\Gamma$-point. Two of them belong to the top and two of them to the bottom layer,  i.e., there is one Dirac cone per surface layer. Therefore HgS is a {\it strong} TI.  The calculated weights of these bands clearly show that they are surface bands with predominantly $3p$ character from the sulfur atoms in the  sub-surface atomic layer. It is remarkable that these surface states decay rather slowly into the bulk (see Fig.~\ref{f4} b), which explains the remaining (tiny) gap for finite layer thickness. 

The two resulting Dirac cones are almost degenerate in the direction $\bar{\Gamma}$-$\bar{X}$ but
in the direction $\bar{\Gamma}$-$\bar{M}$ they split up because of the broken fourfold rotoinversion symmetry at the surface. The direction $\bar{\Gamma}$-$\bar{M}$ corresponds to the diagonal 
$k_{x'}=(k_{x}+k_{y})/\sqrt{2}$ which is identical to the other diagonal $k_{y'}=(k_{x}-k_{y})/\sqrt{2}$ in the tetragonal, rotoinversion symmetric electronic structure of the slab (space group 111), but not at either of the surfaces. In fact, the Dirac cone with the large dispersion along $k_{x'}$ belongs to the top surface and the other one with two nearly flat bands above and below the crossing point corresponds 
to the bottom surface. Along the other diagonal, these roles are interchanged. Indeed, considering just the top Hg-layer and the sulfur atoms in the sub-surface layer, one observes a chain-like structure along one of the diagonals (see Fig.~\ref{f2}).  As will become more clear from the continuum model below, the anisotropy can effectively be described by a crystal field splitting of $p$-orbitals that build the Dirac cone of the surface.

To illustrate our band structure results we construct a simple two-dimensional (2D) model that describes the anisotropic "Dirac cone''. This model displays essential features of the slab band structure in the neighborhood of the $\Gamma$-point. It uses two perpendicular $p$-like orbitals with different on-site energies $\epsilon_{x'}$ and $\epsilon_{y'}$ (crystal field splitting) and directed along the two diagonals of the elementary cell $x'=(x+y)/\sqrt{2}$ and $y'=(x-y)/\sqrt{2}$. The crystal field splitting accounts for the fact that the fourfold rotoinversion symmetry is broken at the surface. The two $p$-like orbitals will be denoted as $|p_{x'} s \rangle$ (for the two spin directions  $s=\uparrow$ or $\downarrow$) and $|p_{y'} s \rangle$. The $p$-like orbitals are no eigenstates of the local SO-coupling operator $\lambda \hat{\bf s} \hat{\bf l}$ but the corresponding coupling elements are easily calculated. The SO-coupling gives also rise to spin-flip hopping in both directions $\alpha k_{x'} \sigma_2$ and $\alpha k_{y'} \sigma_1$. Finally, there are the known tight-binding hoping elements corresponding to $\sigma$-bonds (strength $A$) and $\pi$-bonds (strength $B$). In the continuum limit this Hamiltonian is cast in the form of the 4 by 4 matrix
 
\begin{equation}
H=\left(
\begin{array}{cc}
 E_{x'}\sigma_0  + \alpha k_{x' } \sigma_2  &  -i (\lambda/2) \sigma_3\\
 i (\lambda/2) \sigma_3  &  E_{y'}\sigma_0  + \alpha k_{y' } \sigma_1  \\
\end{array}
\right) \; ,
\label{e1}
\end{equation}
where $\sigma_0$ and $\sigma_{1,2,3}$ are the identity and three Pauli matrices, respectively, and $E_{x'} = \epsilon_{x'}+Ak_{x'}^2+Bk_{y'}^2$ and $E_{y'} = \epsilon_{y'}+Ak_{y'}^2+Bk_{x'}^2$. 
 
\begin{figure}
\begin{center}
\includegraphics[scale=0.35]{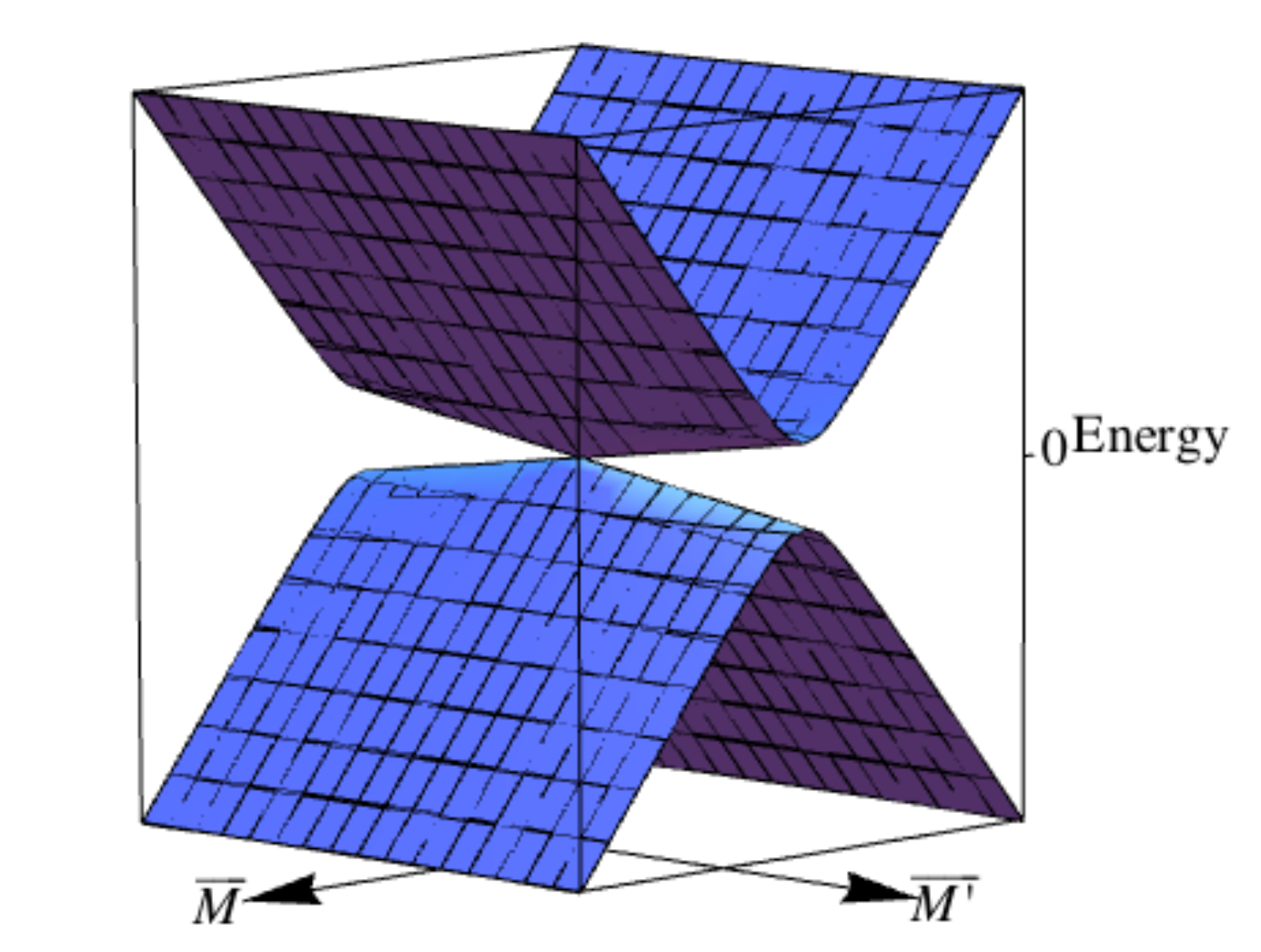}
\includegraphics[scale=0.5]{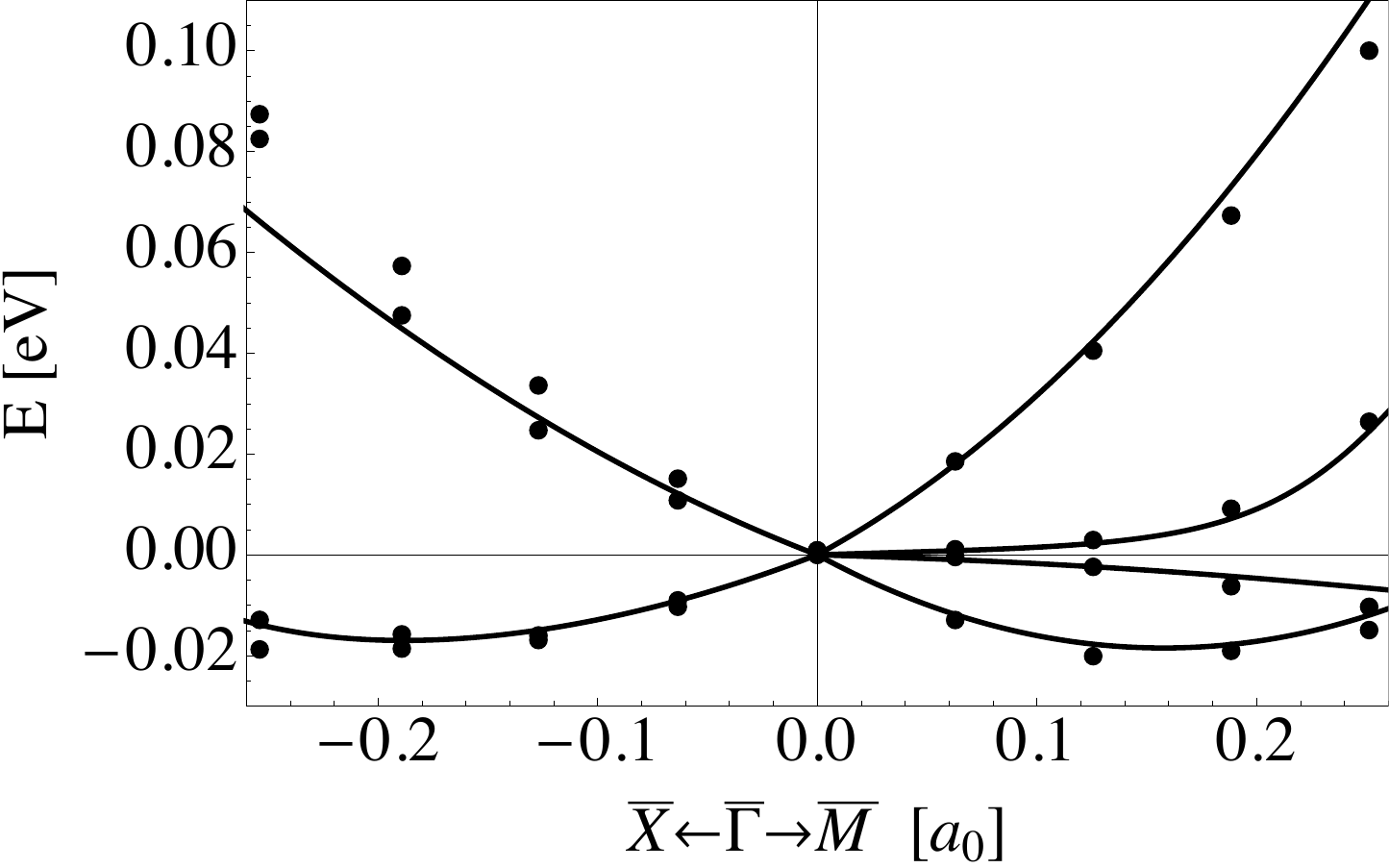}
\end{center}
\caption{(Color) Schematic representation of the anisotropic Dirac cone (top) emerging from the surface bands of an 8-layer slab of HgS (below). The band structure data near the Fermi level are shown as points and the fit with the effective long-wavelength model as lines. The points in the surface Brillouin zone are denoted as $\bar{X}=(\pi/a_0,0)$,  $\bar{M}=(\pi/a_0,\pi/a_0)$, and $\bar{M'}=(\pi/a_0,-\pi/a_0)$ in the $(k_x,k_y)$ coordinate system.} 
\label{f5}
\end{figure}

All momenta $k_{x'}$ and $k_{y'}$ are expressed in units of $1/a_0$ with $a_0=5.85$ \AA. The two upper bands of that model fit the band structure of the 8-layer slab in the neighborhood of the $\Gamma$-point with the following parameters  (see Fig.~\ref{f5}): $\lambda$ = -0.05, $\alpha$ = 0.25, $\Delta E$ = $\epsilon_{x'}$-$\epsilon_{y'}$ =   0.1, $A$ = 0.8, and $B$= -0.1 (all values in eV). For the two relevant surface bands, in the neighborhood of the $\Gamma$-point, the model (\ref{e1}) can be further reduced to the anisotropic, massless Dirac Hamiltonian $H=\alpha_{x'} k_{x'} \sigma_1 + \alpha_{y'} k_{y'} \sigma_2$, with $\alpha_{x'} = \alpha \cos^2 \gamma$, $\alpha_{y'} = \alpha \sin^2 \gamma$, and $\tan (2 \gamma)$ = $|\lambda | / \Delta E$. From our fit we find the anisotropy of the Fermi velocities to be $v_{x'}/v_{y'}$ = $1/\tan^2 \gamma$ $\approx$ 18, which is very large indeed. An additional refinement of this number for the anisotropy ratio requires further density-functional calculations on thicker slabs. If we neglect in the model description the crystal-field splitting, the Dirac cone becomes again isotropic, as it is in for instance Bi$_2$Se$_3$.

To summarize, our density-functional calculations identify metacinnabar, $\beta$-HgS, to be a topological insulator. A detailed study of the (001) surface reveals that its Dirac cone is very anisotropic. The velocity anisotropy $v_{x'}/v_{y'} $ for an eight unit cell slab is $\sim$18. This strong anisotropy
of the topological electronic surface states arises from a broken fourfold rotoinversion symmetry and sharply contrasts the perfectly isotropic bulk band structure. From this observation we expect instead an {\it isotropic} Dirac cone at the (111) surface since the threefold rotation axis along the (111) direction of the bulk remains unbroken at the (111) surface. These predictions can directly be tested by experimental techniques that probe surfaces states, for instance by high precision photo-emission experiments or STM. In particular spin and charge transport at the (001) surface will be very sensitive to the surface anisotropy of the electronic structure that is absent in bulk.

R.H. wishes to thank A. Verga for discussions and the CNRS (PICS project no. 4767) for financial support. JvdB thanks C. Ortix and  J. Venderbos for valuable discussions.

\end{document}